# An Optimization-Based Generative Model of Power Laws Using a New Information Theory Based Metric

A. M. Khalili[1*]
[1]School of Creative Arts and Engineering, Staffordshire University, United Kingdom.
*Correspondence: a.m.khalili@outlook.com.

**Abstract** Power laws have been observed in a wide range of phenomena in fields as diverse as computer science, biology, and physics. Although there are many strong candidate mechanisms that might give rise to power law distributions, no single mechanism has a satisfactory explanation of the origin of power laws in all these phenomena at the same time. In this paper, we propose an optimization-based mechanism to explain power law distributions, where the function that the optimization process is seeking to optimize is derived mathematically, then the behavior and interpretation of this function are analyzed. The derived function shows some similarity to the entropy function in representing order and randomness; however, it also represents the energy, where the optimization process is seeking to maximize the number of elements at the tail of the distribution constrained by the total amount of the energy. The results show a matching between the output of the optimization process and the power law distribution.

## INTRODUCTION

Power laws [1, 2, 3, 4] appear in a wide range of phenomena, they have been observed in city populations [5], the annual incomes of people [6], the number of citations of papers [7], the sales of music and books [8, 9], the number of visits of web pages [10], the magnitudes of earthquakes [11], the numbers of species in biological populations [12], and many other phenomena [13-19]. Fig.1 shows some examples of power laws in different domains.

    There are many candidate mechanisms that might give rise to power law distributions, One of the most accepted mechanisms for producing power laws is the Yule process [20], a rich get richer mechanism where the most cited paper get more citations in proportion to the citations it already has. Yule showed that this mechanism generates the Yule distribution, which has a power law in its tail. This mechanism has been used in explaining paper citations [21, 22], city sizes [23], and links to websites on the internet [24, 25].

    Self-organized criticality [26] has also been proposed as a possible candidate mechanism to generate power law distributions. This mechanism has been used in explaining biological evolution [27] and earthquakes [28, 29].

    Optimization-based mechanisms are also strong candidates. Mandelbrot [30] developed a mechanism to derive power law distributions based on information theory. The mechanism is based on maximizing the amount of information while minimizing the cost of sending that information, where short words cost less energy than long words. Other recent optimization-based mechanisms have been proposed in [31-35]. These mechanisms have been used in explaining the appearance of power laws in the Internet and in human languages.

    In this paper, we propose an optimization-based mechanism, where a new information theory based metric is used to derive power laws. The new metric was derived mathematically from the power law equation, the metric shows some similarity to the entropy function in representing order and randomness; however, it also represents the energy. The results show a matching between the output of the optimization process and the output of the power law.

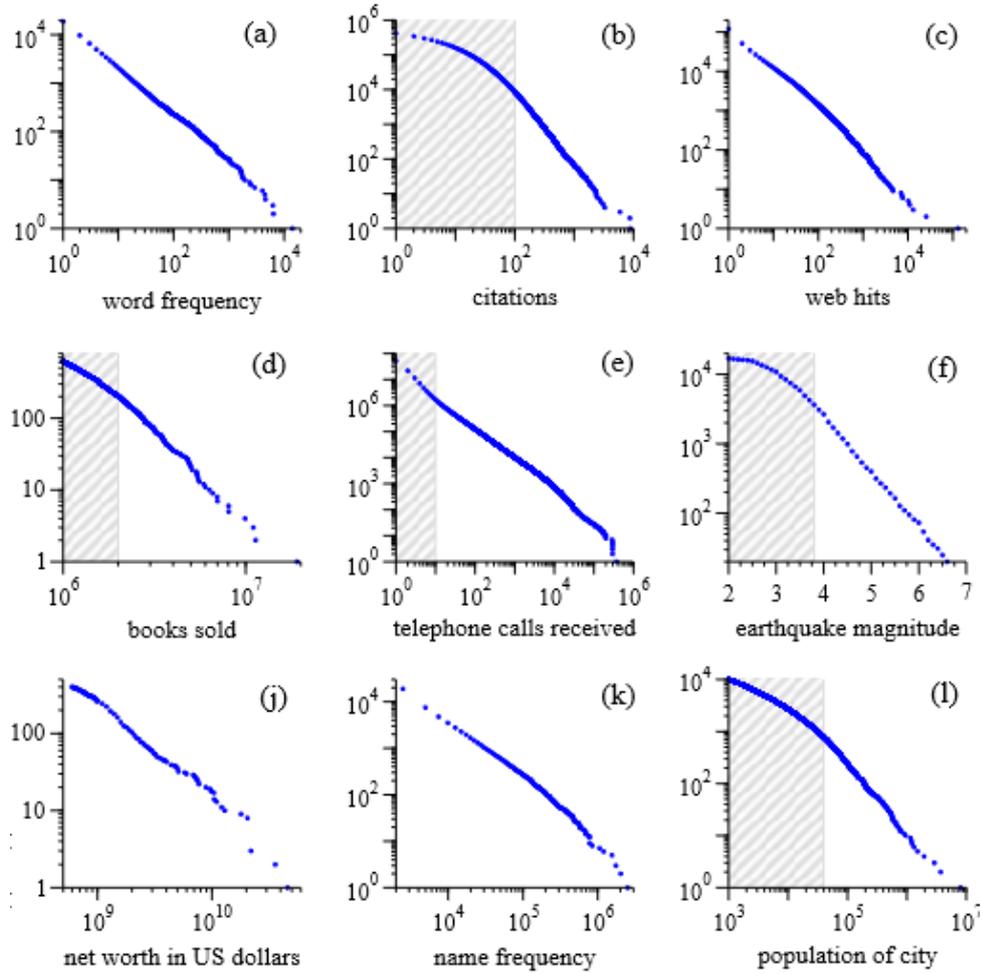

Fig. 1. Some examples of power laws across different domains [1].

**Proposed Approach**

We will assume that power laws are the results of an optimization process, to figure out what the function they seeking to optimize, we will mathematically derive this function then see what it represents, i.e. we will perform 'reverse engineering' to find this function. We will use statistical mechanical terms such as the energy to analyze the behavior of the function; for instance, the elements at $x_1$ will have lower energy than the elements at $x_2$. N is the total number of elements, i.e. the number of papers, cities, etc. The total amount of the energy is related to T, which is the total number of citations, population, wealth, etc. If we begin with the equation of the power law which is given by (1)

$$p(x) = C\, x^{-\alpha} \qquad (1)$$

Where p(x) is the probability of elements at x, C and $\alpha$ are constants. The equation can be written as (2)

$$p(x)x^{\alpha} = C \qquad (2)$$

By converting the probability to the number of elements

$$n(x)x^{\alpha} = CN \qquad (3)$$

Where N is the total number of elements and n(x) is the number of elements at x. By integrating the above function we get

$$n(x)^2 x^\alpha = 2CNn(x) + D \qquad (4)$$

To represent the whole distribution we will take the sum of the terms

$$\sum_i n(x_i)^2 x_i^\alpha = 2CN \sum_i n(x_i) + \sum_i D \qquad (5)$$

The first term represents the function that the power law is seeking to optimize, while the second term is a constraint stating that the total number of elements should be constant. The problem then can be written as the following optimization process

$$Min \; \sum_i n(x_i)^2 x_i^\alpha \qquad (6)$$

Subject to

$$\sum_i n(x_i) = Constant \qquad (7)$$

Minimizing the function in (6) will results in a power law distribution. The value of $\alpha$ can be determined by knowing T, which is given by

$$T = \sum_i n(x_i) x_i \qquad (8)$$

Using (3), the number of elements can be written as

$$n(x) = CN\, x^{-\alpha} \qquad (9)$$

Substituting (9) in (8) we get

$$T = CN \sum_i x_i^{-\alpha+1} \qquad (10)$$

Since T, N, C, $x_1$ $x_2$ ....... $x_L$ are known, then the value of $\alpha$ can be calculated.

The above derivation is very similar to the derivation of the Maxwell-Boltzmann distribution [36, 37, 38], which has recently found applications in economics [39, 40], social behaviors [41] and aesthetic [42]. However, the Maxwell-Boltzmann distribution seeks to maximize the entropy. Boltzmann showed that this distribution is the most probable distribution and it will result by maximizing the multiplicity (which is equivalent to maximizing the entropy), where the number of particles should be constant (11), the energy levels that the particles can take should be constant (12), and the total energy should be constant (13). The multiplicity is described by (14), and the entropy is described by (15).

$$\sum_i n_i = Constant \; (11)$$

$$x_1, x_2, \ldots, x_n \; Constant \; (12)$$

$$Energy = \sum_i n_i x_i = Constant \; (13)$$

$$\Omega = \frac{N!}{n_1! n_2! \ldots n_n!} \; (14)$$

$$Entropy = \log(\Omega) \; (15)$$

Where N is the total number of particles, $n_i$ is the number of particles at the $x_i$ energy level.

Minimizing the function in (6) will result in increasing the number of elements at high $x_i$ values; however, this is constraint by T, i.e. the total citations, the total population, the total wealth, etc. Since the function in (6) represents both the energy and the distribution of the elements, we will try to separate them. The function in (6) has some similarity to the entropy function in representing order and randomness. However, the function also represents the energy; for instance, the entropy will have the same value if all the elements are at $x_1$ or all the elements are at $x_2$, while the new function will produce higher value when all the elements are at $x_2$. The $n(x_i)^2$ factor in (6) has a spreading effect; it will be minimum when all the elements are equally distributed among all levels $x_1\ x_2\ \ldots\ x_L$. On the other hand, the $x_i^\alpha$ factor will be minimum when all the elements are at $x_1$. Which means that the function represents two forces, the first one is trying to spread the elements to higher levels, while the other is trying to keep the elements at the minimum energy level.

**Results**

Minimizing the function in (6) will drive the elements to reach higher $x_i$ values. To analyze the behavior of the function in (6), we will assume that the number of elements is N = 100000, the number of levels is L = 10 and α = 3. We assume that at the first time step, all elements are at $x_1$. Now we will allow one element at each time step to move from one level to the next level, such as moving from $x_1$ to $x_2$. The only constraint on the movement of the elements is that they should move to the level that minimizes the function in (6), i.e. from all possible movements, the movement that minimizes the function will be chosen. When the minimum value was reached, no movement minimized the function anymore. The resulted distribution at the minimum value matches with the distribution of the power law. The first following line represents the power law distribution, and the second line represents the distribution resulting from the optimization process. Fig. 2 shows how the distribution is changing when increasing T.

| 83505 | 10438 | 3093 | 1305 | 668 | 387 | 243 | 163 | 115 | 83 |
|---|---|---|---|---|---|---|---|---|---|
| 83509 | 10439 | 3093 | 1305 | 668 | 386 | 243 | 162 | 113 | 82 |

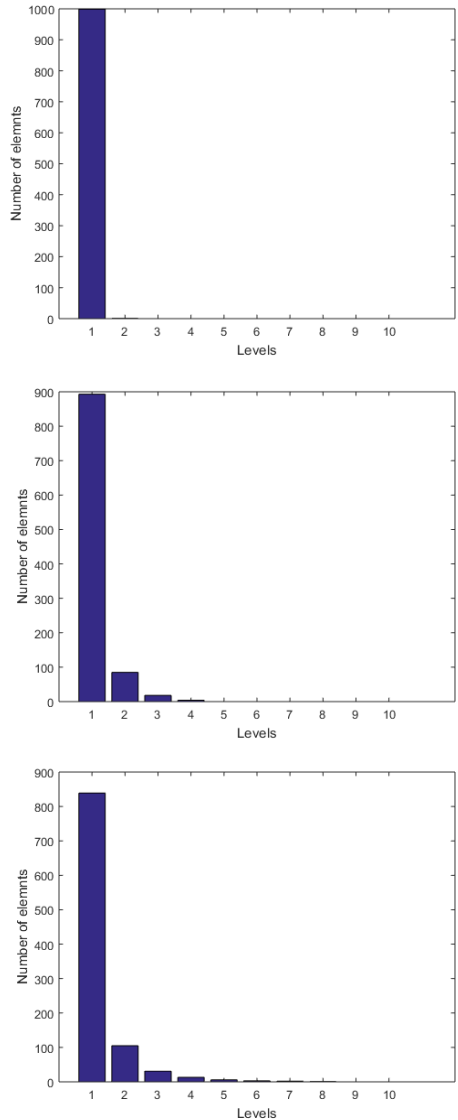

Fig. 2. The behavior of the function in (6) when increasing T.

The next important question would be why the function described by (6) is used in the optimization process. If we perform the same analysis for the entropy function, after 70 time steps we get

```
249  43   7   1   0   0   0   0   0   0
255  30   9   3   2   1   0   0   0   0
```

The first line is the result of maximizing the entropy, while the second line is the result of minimizing the function in (6). The entropy function is trying to flatten the distribution at the low levels before moving more elements to higher levels, while the other function is focusing on increasing the number of elements at higher levels. This means that there are weaker interactions between the elements and the elements have more freedom to move to higher levels. While in the entropy case, there is a stronger coupling between the elements, i.e. the function in (6) has more ability to move elements to higher levels than the entropy function when the same energy or cost is used.

## Discussion

The power law distribution represents the optimal point between two forces; the first one represents the desire of the element to move to higher levels and second force tries to constrain the spreading of the elements by the amount of the available energy in the system, where the movement of each element requires more cost or energy, and the element that can spend this energy will move to the higher level. The function in (6) represents the optimization process independently from the actual reasons that derive the elements in both directions.

In networks context for instance, the reason behind the force that tries to keep the elements localized at low levels is the total cost, i.e. we want to minimize the number of links of each node to minimize the cost. The reason behind the force that tries to spread the elements and increase the number of links of each node is to minimize the distance to other nodes, this will minimize the delay, the congestion, and the energy consumptions; however, this will increase the cost.

Similar reasons could be identified for many other phenomena. In general, there are two groups of forces, the first one tries to increase the number of elements at higher levels, and the second one tries to constrain this increment by the available cost or energy.

The mechanism has shown a link between information theory and power laws. The deeper understanding of this link is to be further investigated in future work. The proposed approach has also presented a new information theory based function. Unlike the entropy, the function represents both the energy and the distribution of the elements.

The derivation of the mechanism is very similar to the derivation of the Maxwell-Boltzmann distribution. Maximizing the entropy will result in the Maxwell-Boltzmann distribution, while minimizing the derived function will result in power law distributions. It is interesting to see whether this function will help in providing new perspectives to other phenomena where power law distributions are also showing up.

## Conclusion

This paper has proposed an optimization-based mechanism to generate power law distributions, where the function that the optimization process is seeking to optimize is derived mathematically. The derived function shows some similarity to the entropy function in representing order and randomness; however, it also represents the energy. The results have shown that the proposed approach was able to match the power law distribution.